\documentclass{article}
\usepackage{graphicx}
\usepackage{amsmath}

\input{tcilatex}

\begin{document}

\begin{center}
{\LARGE VLHC Based ep Colliders: e-ring versus e-linac}

\bigskip

Y. Islamzade$^{1)}$,{\large \ }H. Karadeniz$^{2)}$, S. Sultansoy$^{1,3)}$
\end{center}

$^{1)}$Dept. of Physics, Faculty of Arts and Sciences, Gazi University,
06500 Teknikokullar, Ankara, Turkey.

$^{2)}$Ankara Nuclear Research and Training Center, 06100 Besevler, Ankara,
Turkey.

$^{3)}$Institute of Physics, Academy of Sciences, H. Cavid Ave. 33, Baku,
Azerbaijan.

\bigskip

\bigskip

\begin{center}
\textbf{Abstract}
\end{center}

Main parameters of \ a Linac$\ast $VLHC based ep collider are estimated and
compared with recently suggested ep collider in the VLHC tunnel.

\bigskip

\textbf{I. Introduction}

\bigskip

The CERN Large Hadron Collider (LHC) and 50+50 TeV hadron collider (VLHC),
sub-TeV (TESLA, JLC/NLC) and multi-TeV (CLIC) electron-positron colliders
and multi-TeV energy muon collider (HEMC) are considered as energy frontiers
for the 21st century high energy physics research [1]. On the other hand,
the importance of lepton-hadron colliders is a matter upon which general
agreement is made. Indeed, the quark-parton structure of the matter was
discovered in lepton-hadron collisions. The first and unique operating
electron-proton collider is HERA with $\sqrt{\text{s}}\approx 300$ GeV and L$%
\approx 10^{31}$ cm$^{-2}$s$^{-1}$. Nowadays, an addition of e-ring or
e-linac to RHIC (eRHIC) is intensively discussed [2]. There are two
proposals in achieving TeV scale, namely, LEP$\ast $LHC [3] and THERA [4],
that can be taken seriously. Concerning multi-TeV scale in lepton-hadron
collisions, the Linac$\ast $LHC based ep, $\gamma $p, eA and $\gamma $A
colliders [5] are of the most realistic feature in contrast to speculative $%
\mu $p collider proposals.

There are a number of papers devoted to possible ep collider based on VLHC.
Two options are evaluated [6]: ep collisions in VLHC booster ($\sqrt{\text{s}%
}=1$ TeV and L$=2.6$x$10^{32}$ cm$^{-2}$s$^{-1}$) and ep collisions in VLHC
ring ($\sqrt{\text{s}}=6$ TeV and L$=1.4$x$10^{32}$ cm$^{-2}$s$^{-1}$). The
first option is left untouched in this paper because of LEP$\ast $LHC and
THERA covering the same region. For the second option, where the
construction of E$_{\text{e}}$=180 GeV e-ring in the VLHC tunnel is planned,
we have two objections:

1. Linac$\ast $LHC with $\sqrt{\text{s}}=5$ TeV and L$=10^{33}$ cm$^{-2}$s$%
^{-1}$can be realized earlier and will give \ opportunity to an additional $%
\gamma $p, eA and $\gamma $A options.

2. Instead of constructing a 530 km e-ring it is more wise to construct a 10
km e-linac with the same parameters.

\vspace{1cm}

\bigskip

\bigskip

\bigskip

\bigskip

\textbf{II. Linac}$\ast $\textbf{VLHC based ep colliders}

\bigskip

Parameters of VLHC proton beam and electron beam in VLHC tunnel are
presented in the second and third columns of \ Table 1. For e-linac options
we consider TESLA [7] and the JLC/NLC [8] beams with 250 GeV. Corresponding
parameters are given in last two columns of the Table.

Luminosity of ep collisions is calculated using:

\begin{equation*}
L=\frac{n_{e}n_{p}}{4\pi \varepsilon _{t}\beta ^{\ast }}f_{coll}
\end{equation*}
where n$_{e}$ and n$_{p}$ are the numbers of electrons and protons in the
corresponding bunches, $\varepsilon _{t}$ is the transverse emittance of
proton beam which is equal to $\varepsilon _{N}/\gamma _{p}$ ($\varepsilon
_{N}$ is the normalized emittance and $\gamma _{p}$ is the Lorentz factor
for proton beam), $\beta ^{\ast }$ is amplitude function at the interaction
region and $f_{coll}$ is collision frequency which is equal to $n_{b}f_{rep}$
$(n_{b}$ stands for number of bunches in electron pulse, $f_{rep}$ is linacs
repetition frequency) for linac-ring colliders and to $2\pi R/ck_{b}$ (2$\pi 
$R is VHLC circumference, c is the speed of light, k$_{b}$ is number of
proton and electron bunches in ring) for ring-ring colliders.

In the case of linac-ring type colliders bunch spacing of electron linac
should be adjusted to match with bunch spacing of proton ring. For the
''TESLA'' option, these spacing are of the same order, so the adjustment is
not a particular problem. However for the ''JLC/NLC'' option the bunch
spacing of electron beam (2.8 ns) is smaller than proton spacing in VLHC by
orders. For this reason we consider the apprapriate upgrade of JLC/NLC
parameters, namely, number of bunches per pulse is lowered by an order (95$%
\rightarrow $10) with simultaneous increase in number of electrons per bunch
(10$^{10}\rightarrow $10$^{11}$). At the same time number of bunches in VLHC
proton ring adjusted from 6000 to 60000.

For the luminosity evaluation we take $\varepsilon _{N}$ and $\beta ^{\ast }$
to be 10$^{-\text{6}}$mrad and 10 cm, respectively. With these values we
obtain $L_{ep}\approx 3$x$10^{32}$ cm$^{-2}$s$^{-1}$ for ''TESLA''$\ast $%
VLHC and $L_{ep}\approx 2$x$10^{32}$ cm$^{-2}$s$^{-1}$ for ''JLC/NLC''$\ast $%
VLHC.

In principle, the luminosity can be increased further using ''dynamic
focusing'' [9] and/or cooling of proton beams. Applying both of these
methods together, at least an order higher luminosity can be achieved.

\bigskip \bigskip

\textbf{III. Conclusion}

\bigskip

A high energy lepton-hadron colliders will provide data crucial to
exploration of the ultimate constituents of matter. Concerning the Standard
Model, these machines will essentially enlarge our understanding of QCD,
especially in the domain of small x$_{\text{B }}$and large Q$^{2}$ which is
vital for confinement problem. The data will be used to interpret results
from future high energy hadron colliders, also. In the field of BSM physics,
there are a lot of phenomena that can be best investigated in lepton-hadron
collisions, such as, leptoquarks and leptogluons, excited leptons ($\nu
^{\ast }$, e$^{\ast }$), sneutrino and sleptons etc. The $\gamma $p option
of future linac-ring type lepton-hadron colliders, will essentially enlarge
their physics research potential. Indeed, photon-hadron colliders are the
best machines for investigation of small x$_{\text{g}}$ region via pair
production of heavy quarks, associated production of sneutrino and squarks,
resonant production of excited quarks etc (For more details on the subject
see review [10]).

As a result, future lepton-hadron colliders (see Table 2) will be
complementary to future hadron and lepton colliders. In near term future
different upgrades of the HERA (luminosity, polarized protons, nuclear
beams) and eRHIC machines promise valuable results. Concerning TeV scale,
the THERA (TESLA$\ast $HERA) can be the first machine and Linac$\ast $LHC
should be considered seriously as the next step. As mentioned above, an
additonal options of THERA and Linac$\ast $LHC, namely, $\gamma $p, eA and $%
\gamma $A colliders extend considerably their physics research potential. As
for far term future, Linac$\ast $VLHC (rather than e-ring in VLHC tunnel)
requires more serious consideration.

\bigskip

\textbf{Acknowledgements}

\bigskip

We are grateful to A. K. Ciftci and O. Yavas for useful discussions. This
work is supported in part by Turkish Plannig Organization (DPT) under the
Grant No 2002K120250.

\bigskip

\textbf{References}

\bigskip

[1] www.snowmass2001.org

[2] I. Ben-Zvi et al., Nucl. Instum. Meth. A463, 94 (2001).

[3] J. R. Ellis, E. Keil and G. Rolandi, CERN-EP-98-03 (1998).

[4] www.rfh.de/thera;\ O. Yavas, A.K. Ciftci and S. Sultansoy, Proc. of the
\ \ \ \ 

\ \ \ \ \ 7 th European Accelerator Conference (EPAC2000), p.388,

\ \ \ \ \ hep-ex/0004013.

[5] O. Yavas, A.K. Ciftci and S. Sultansoy, Proc. of the 7 th European

\ \ \ \ \ Accelerator Conference (EPAC2000), p.391, hep-ex/0004016.

[6] J. Norem and T. Sen, FERMILAB-PUB-99/347 (1999).

[7] TESLA Technical Design Report, DESY 2001-011 and ECFA 2001-209

\ \ \ \ (2001).

[8] KEK Report 2000-7 and SLAC-R-559 (2000).

[9] R. Brinkmann and M. Dohlus, DESY-M-95-11 (1995); R. Brinkmann,

\ \ \ \ Turk. J. of Phys. 22, 661 (1998).

[10] S. Sultansoy, Turk. J. of Phys. 22, 575 (1998);

\ \ \ \ \ \ S. Sultansoy,DESY-99-159 (1999), hep-ph/9911417;

\ \ \ \ \ \ S. Sultansoy, A.K. Ciftci,\ E.Recepoglu and O. Yavas,
hep-ex/0106082.

\bigskip \vspace{1cm}\medskip \bigskip

\bigskip

\bigskip

\bigskip

\bigskip

\bigskip

Table 1. Main parameters of the VLHC based ep colliders (TESLA parameters
are given for THERA option).\vspace{0.3cm}

\begin{tabular}{|l|l|l|l|l|}
\hline\hline
& VLHC & e-ring & TESLA & JLC/NLC \\ \hline
Beam energy, TeV & \multicolumn{1}{|c|}{50} & \multicolumn{1}{|c|}{0.18} & 
\multicolumn{1}{|c|}{0.25} & \multicolumn{1}{|c|}{0.25} \\ \hline
Circumference or length, km & \multicolumn{1}{|c|}{531} & 
\multicolumn{1}{|c|}{531} & \multicolumn{1}{|c|}{10} & \multicolumn{1}{|c|}{
4.3} \\ \hline
Luminosity for ep, 10$^{32}$cm$^{-2}$s$^{-1}$ & \multicolumn{1}{|c|}{} & 
\multicolumn{1}{|c|}{1.4} & \multicolumn{1}{|c|}{3} & \multicolumn{1}{|c|}{$%
\rightarrow $2} \\ \hline
Center of mass energy, TeV & \multicolumn{1}{|c|}{} & \multicolumn{1}{|c|}{
6.0} & \multicolumn{1}{|c|}{7.07} & \multicolumn{1}{|c|}{7.07} \\ \hline
Number of bunches per ring & \multicolumn{1}{|c|}{6000} & 
\multicolumn{1}{|c|}{6000} & \multicolumn{1}{|c|}{-} & \multicolumn{1}{|c|}{-
} \\ \hline
Particles per bunch, 10$^{10}$ & \multicolumn{1}{|c|}{12.5} & 
\multicolumn{1}{|c|}{10.1} & \multicolumn{1}{|c|}{2} & \multicolumn{1}{|c|}{1%
$\rightarrow $10} \\ \hline
Number of bunches per pulse & \multicolumn{1}{|c|}{-} & \multicolumn{1}{|c|}{
-} & \multicolumn{1}{|c|}{5600} & \multicolumn{1}{|c|}{95$\rightarrow $10}
\\ \hline
Pulse frequency, Hz & \multicolumn{1}{|c|}{-} & \multicolumn{1}{|c|}{-} & 
\multicolumn{1}{|c|}{5} & \multicolumn{1}{|c|}{150} \\ \hline
\end{tabular}

\begin{center}
\bigskip

\bigskip
\end{center}

Table 2. Future TeV scale lepton-hadron colliders.

\bigskip 
\begin{tabular}{|c|ccc|rc|rl|}
\hline\hline
&  & THERA &  & Linac $\ast $ & LHC & Linac $\ast $ & VLHC \\ \hline
& 1 & \multicolumn{1}{|c}{2} & \multicolumn{1}{|c|}{3} & \multicolumn{1}{|c}{
1} & \multicolumn{1}{|c|}{2} & \multicolumn{1}{|c}{1} & \multicolumn{1}{|c|}{
2} \\ \hline
E$_{\text{e}}$, TeV & 0.25 & \multicolumn{1}{|c}{0.5} & \multicolumn{1}{|c|}{
0.8} & \multicolumn{1}{|c}{1} & \multicolumn{1}{|c|}{1} & 
\multicolumn{1}{|c}{0.25} & \multicolumn{1}{|c|}{3} \\ \hline
E$_{\text{p}}$, TeV & 1 & \multicolumn{1}{|c}{0.5} & \multicolumn{1}{|c|}{0.8
} & \multicolumn{1}{|c}{7} & \multicolumn{1}{|c|}{14} & \multicolumn{1}{|c}{
50} & \multicolumn{1}{|c|}{50} \\ \hline
$\sqrt{\text{s}}$, TeV & 1 & \multicolumn{1}{|c}{1} & \multicolumn{1}{|c|}{
1.6} & \multicolumn{1}{|c}{5.29} & \multicolumn{1}{|c|}{7.48} & 
\multicolumn{1}{|c}{7.07} & \multicolumn{1}{|c|}{24.5} \\ \hline
L, 10$^{31}$cm$^{-2}$s$^{-1}$ & 0.4 & \multicolumn{1}{|c}{2.5} & 
\multicolumn{1}{|c|}{1.6} & \multicolumn{1}{|c}{10$\div $100} & 
\multicolumn{1}{|c|}{20$\div $200} & \multicolumn{1}{|c}{20$\div $200} & 
\multicolumn{1}{|c|}{10$\div $100} \\ \hline
\end{tabular}

\end{document}